\documentclass[conference]{IEEEtran}
\IEEEoverridecommandlockouts
\usepackage{cite}
\usepackage[fleqn]{amsmath}
\usepackage{amssymb,amsfonts}
\usepackage{algorithm}
\usepackage{algpseudocode}
\usepackage{graphicx}
\usepackage{textcomp}
\usepackage{cleveref}
\usepackage{url}
\usepackage{xcolor}

\def\BibTeX{{\rm B\kern-.05em{\sc i\kern-.025em b}\kern-.08em
    T\kern-.1667em\lower.7ex\hbox{E}\kern-.125emX}}
\usepackage{acronym}
\usepackage{enumitem}

\begin{document}

\newacro{QoS}[QoS]{Quality of Service}
\newacro{BS}[BS]{base station}
\newacro{RB}[RB]{resource block}
\newacro{RBs}[RBs]{resource blocks}
\newacro{OFDMA}[OFDMA]{Orthogonal Frequency Division Multiple Access}
\newacro{CL}[CL]{Capacity Limited}
\newacro{URLLC}[URLLC]{Ultra Reliable Low Latency Communication}
\newacro{TS}[TS]{Time Sensitive}
\newacro{5G}[5G]{5th Generation}
\newacro{NS}[NS]{network slicing}
\newacro{RRA}[RRA]{radio resource allocation}
\newacro{AGV}[AGV]{Automated Guided Vehicle}
\newacro{5GACIA}[5GACIA]{5G Alliance for Connected Industries and Automation}
\newacro{PLC}[PLC]{programmable logic controller}
\newacro{InF}[InF]{Indoor Factory}
\newacro{IoT}[IoT]{Internet of Things}
\newacro{IIoT}[IIoT]{Industrial IoT}
\newacro{RRM}[RRM]{radio resource management}
\newacro{3GPP}[3GPP]{3rd Generation Partnership Project}
\newacro{LOS}[LOS]{Line-of-sight}
\newacro{NLOS}[NLOS]{Nonline-of-sight}
\newacro{ML}[ML]{Machine Learning}
\newacro{AI}[AI]{Artificial Intelligence}
\newacro{DRL}[DRL]{Deep Reinforcement Learning}
\newacro{E2E}[E2E]{end-2-end}
\newacro{SFC}[SFC]{service function chains}
\newacro{VNE}[VNE]{Virtual Network Embedding}
\newacro{HMI}[HMI]{Human Machine Interface}

\crefname{figure}{Fig.}{Figs.}
\Crefname{figure}{Figure}{Figures}

\title{Slicing for Dense Smart Factory Network: Current State, Scenarios, Challenges and Expectations 
}

\author{\IEEEauthorblockN{Regina Ochonu, Josep Vidal}
\IEEEauthorblockA{\textit{Dept. of Signal Theory and Communications, Universitat Politècnica de Catalunya, Barcelona, Spain} \\
\{regina.ochonu, josep.vidal\}@upc.edu}}

\maketitle

\begin{abstract}
In the era of Industry 4.0, smart factories have emerged as a paradigm shift, redefining manufacturing with the integration of advanced digital technologies. Central to this transformation is the deployment of 5G networks, offering unprecedented levels of connectivity, speed, reliability, and ultra-low latency. Among the revolutionary features of 5G is network slicing, a technology that offers enhanced capabilities through the customization of network resources by allowing multiple logical networks (or "slices") to run on top of a shared physical infrastructure. This capability is particularly crucial in the densely packed and highly dynamic environment of smart factories, where diverse applications—from robotic automation to real-time analytics—demand varying network requirements. In this paper, we present a comprehensive overview of the integration of slicing in smart factory networks, emphasizing its critical role in enhancing operational efficiency and supporting the diverse requirements of future manufacturing processes. We elaborate on the recent advances, and technical scenarios, including indoor factory propagation conditions, traffic characteristics, system requirements, slice-aware radio resource management, network elements, enabling technologies and current standardisation efforts. Additionally, we identify open research challenges as well as key technical issues stifling deployments. Finally, we speculate on the future trajectory of slicing-enabled smart factories, emphasizing the need for continuous adaptation to emerging technologies.

\end{abstract}


\section{Introduction}
In the evolving landscape of global manufacturing, the demand for advanced, time-sensitive, smart, adaptable, and automated industrial control systems is at an all-time high. And the limitations of traditional wired network infrastructures in current factories have become increasingly apparent. These wired systems, while reliable, lack the flexibility and ease of reconfiguration necessary for a smart factory environment. A Smart Factory is an advanced manufacturing solution designed to adapt and respond effectively to complex and evolving production scenarios \cite{b1}. This solution has two main facets: on one hand, it involves the fusion of software, hardware, and mechanical systems to advance automation, aiming to optimize manufacturing processes, and thereby reducing labour and resource waste. On the other hand, it emphasizes collaboration among a variety of industrial and non-industrial stakeholders, with the 'smartness' stemming from the creation of a dynamic and responsive organizational network\cite{b1}\cite{b2}.

\begin{figure}[htbp]
\centerline{\includegraphics[width=0.5\textwidth]{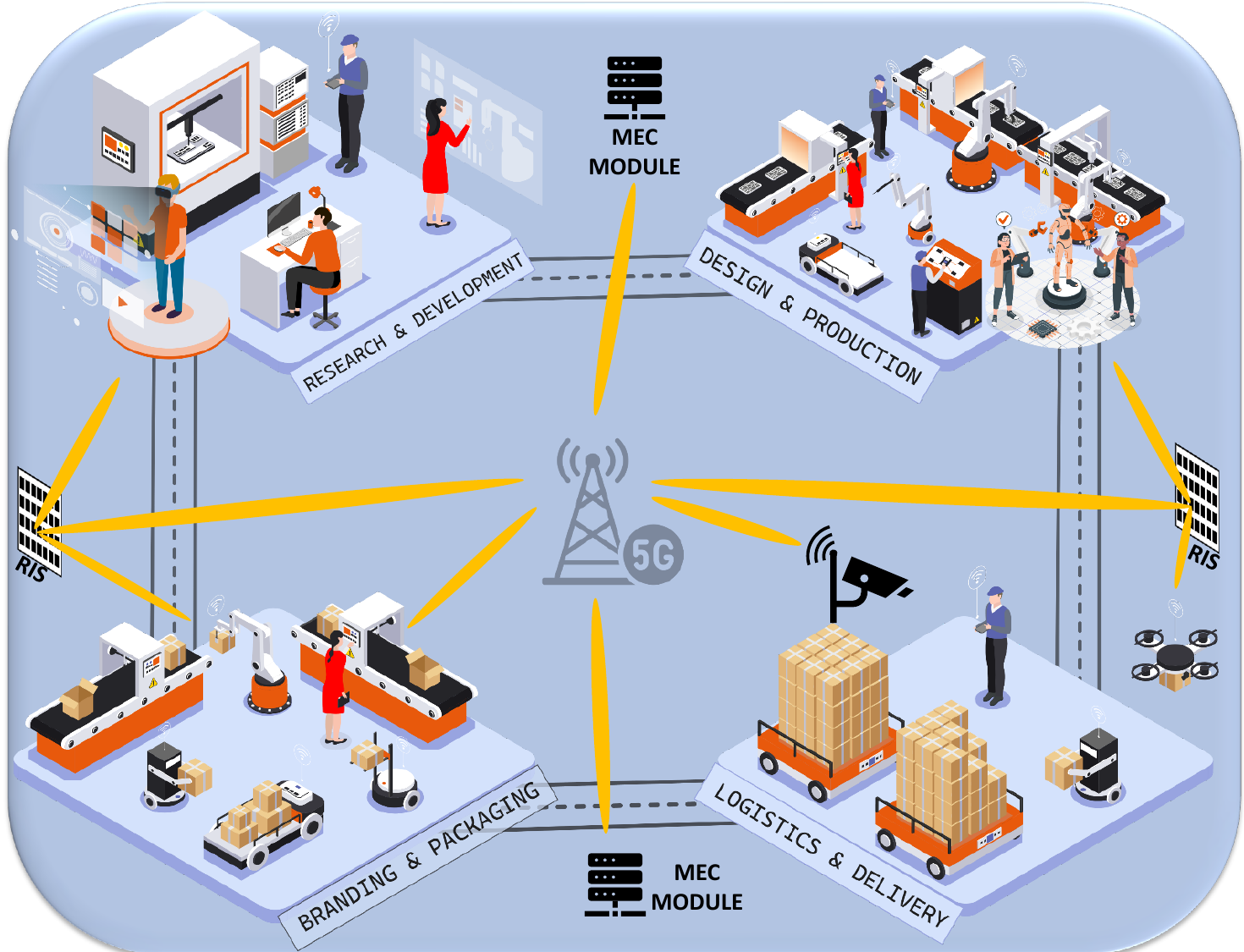}}
\caption{5G Enabled Smart Factory with an ultra-dense network}
\label{fig:5Genabled_SmactFact}
\end{figure}

The emergence of the 5th-generation (5G) network offers more flexible wireless connections that align with the requirements of smart factory networks. 5G capabilities—such as ultra-reliable and low latency, massive device connectivity, enhanced data rates, and availability — will facilitate industrial processes including real-time monitoring, automation, and overall system optimization \cite{b3}. A 5G-enabled smart industry, illustrated in \cref{fig:5Genabled_SmactFact}, is a fully digitized and interconnected production facility, harnessing the capabilities of 5G network to facilitate a cohesive and collaborative ecosystem capable of autonomous operation and decentralized decision-making\cite{b3}. A unique feature of 5G that can significantly enhance the digitization of manufacturing processes is \acf{NS}. 

\ac{NS} is a transformative technology that enables the creation of multiple virtual networks (called slices) over a single physical network infrastructure\cite{b4}. Each slice is an independent end-to-end network that can be customized for the specific requirements of particular applications or services \cite{b5}. The concept of network slicing is not just a technological advancement; it represents a fundamental shift in how industrial wireless networks are designed, managed, and optimized\cite{b6}. By allowing for the customization of network attributes such as latency, bandwidth, and security, network slicing holds the promise of a highly flexible, scalable, and efficient manufacturing ecosystem tailored to the specific needs of diverse industrial applications. 

Despite its potential, implementing network slicing in the context of smart factories presents significant challenges. The dense radio environment of the factory floor, characterized by a high degree of electromagnetic interference and physical obstacles, introduces uncertainties in the wireless channel and unveils unique challenges for network design and management. Furthermore, the mobility of terminals like \ac{AGV}, varied types of network traffic, and the limited availability of radio resources, such as bandwidth, add to the complexity.  Moreover, the critical nature of manufacturing processes requires the network to deliver unparalleled levels of reliability and real-time performance.

In this paper, we provide a holistic overview of network slicing in the smart factory context, highlighting the potential benefits, existing challenges, and future prospects with the evolution towards advanced network technologies like 6G. To our knowledge, no existing survey has primarily focused on slicing for dense smart factory networks. As network slicing is a crucial technology driving smart manufacturing and facilitating the implementation of Industry 4.0, it is essential to explore this topic. The main contributions of this paper are summarised as follows:
\begin{itemize}
    \item An in-depth overview of network slicing within the domain of smart manufacturing, complemented by a review of the current advancements in 5G network slicing applicable to dense smart factory networks.
    \item A comprehensive description of technical scenarios that encompass the system requirements for successful factory deployments, roles of different network elements, characteristics of in-factory traffic, the 3GPP channel model for indoor-factory radio propagation, approaches for slice-aware resource management,  enabling technologies, and the latest efforts towards standardisation.
    \item Highlight existing research gaps in slicing for smart factory networks and offer perspectives on possible future developments.
\end{itemize}  

The remainder of this paper is organized as follows: Section II provides a discussion on 5G support for network slicing along with an overview of slicing in smart factory networks. Section III highlights recent advancements in network slicing for smart factory applications and identifies limitations in these methods. Section IV elaborates on various technical scenarios, while Section V explores open research challenges. The paper concludes with Section VI, offering perspectives on future directions.

\section{5G Support For slicing}
The 5G architecture comprises the 5G Access Network (AN), 5G Core Network, and User Equipment (UE)\cite{b7}. The 5G system will offer optimized support for diverse communication services, varying traffic loads, and different user communities\cite{b8}. As a result, the traditional approach of a 'one size fits all' network does not suit the needs of 5G and beyond wireless technologies. In response to this, the Next Generation Mobile Network (NGMN) Alliance introduced the concept of \ac{NS} in 2015 in \cite{b9}. According to NGMN, the network slicing process is divided into three main layers: the service instance layer, the network slice instance layer, and the resource layer\cite{b10}. Each service instance caters to services from specific sectors or providers, while the network slice instance comprises resources tailored to meet specific service requirements, potentially including multiple sub-network instances that may be isolated or shared. These sub-networks can operate independently or interact and can span across different administrative and technological domains\cite{b11}. This structure allows for the dynamic control and automation of network slices through common resource abstractions and programmable interfaces, effectively accommodating evolving service demands.

Network slice management encompasses four distinct phases\cite{b12}\cite{b13}: 
\begin{enumerate}
\item \textit{Preparation:} in this phase, no Network Slice Instance (NSI) exists and it involves tasks like design and capacity planning. NSI provides the network
characteristics required by a Service Instance. 
\item \textit{Commissioning:} includes the creation and configuration of the NSI to fulfill specific requirements. 
\item \textit{Operation:} covers activation, monitoring, modification based on performance reports or new requirements, and deactivation to halt services.
\item \textit{Decommissioning:} here, non-shared elements are decommissioned and configurations from shared constituents are removed, culminating in the termination of the NSI.
\end{enumerate}
Each phase is critical for the systematic establishment, maintenance, and closure of network slices, ensuring effective management throughout their lifecycle.

\subsection{Slicing for Smart Factory use case}
One of the objectives for the development of 5G was to enhance wireless connectivity across various vertical industries, notably in manufacturing\cite{b14}. To fulfil this objective, 5G is designed to support three fundamental communication frameworks: enhanced mobile broadband (eMBB), massive machine-type communications (mMTC), and ultra-reliable low-latency communications (URLLC). eMBB delivers very high data rates, in the order of several Gb/s, alongside broader network coverage. mMTC enables connectivity for a massive number of IoT devices, up to hundreds of thousands per square kilometre, ensuring extensive coverage and deep indoor penetration. This framework is also engineered to minimize the software and hardware requirements on devices, facilitating energy-efficient operations ideal for IoT ecosystems. Meanwhile, URLLC caters to critical applications that demand very low end-to-end latency, on the millisecond scale, and high reliability and availability, essential for high-stakes environments like smart industries\cite{b8}.

\begin{figure}[htbp]
\centerline{\includegraphics[width=0.47\textwidth]{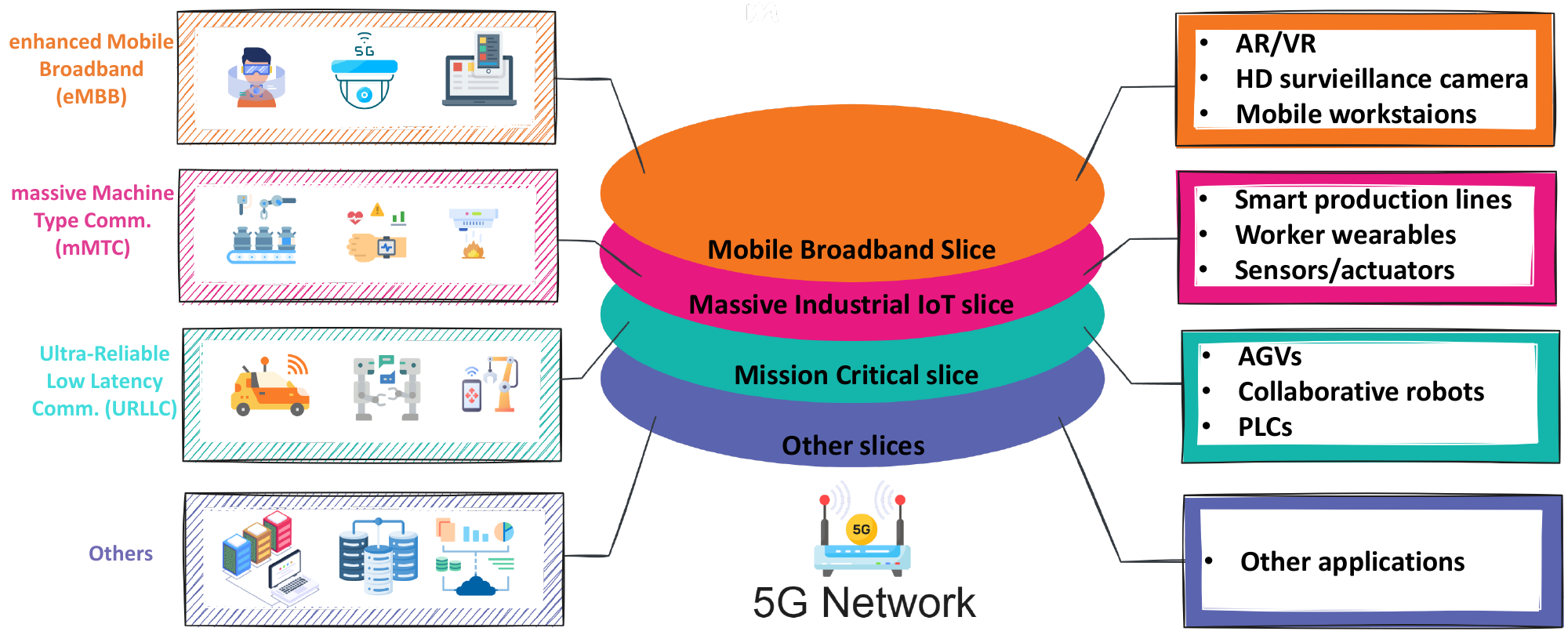}}
\caption{Slicing for Smart Factory Applications}
\label{fig:NS_SmactFact}
\end{figure}

To accommodate these varied communication needs and the complex requirements of anticipated 5G applications, 5G introduces \ac{NS} which allows for the simultaneous, yet isolated, provisioning of diverse services within the same network infrastructure. By employing \ac{NS}, a  smart factory can allocate and optimize its network resources efficiently, ensuring that each application receives the level of service it requires as illustrated in \cref{fig:NS_SmactFact}. This not only improves operational efficiency and productivity but also enhances safety and flexibility across the factory floor.

\section{Recent Advances}
The vision of 5G \ac{NS} relies heavily on the full automation of network operations and resource management, a goal that has seen considerable advancement through the integration of \ac{AI} and \ac{ML} techniques.

Traditionally, network slicing problems are often modeled as either Mixed-integer Linear Programming (MILP) or Nonlinear Integer Programming (NLP) optimization problems due to their ability to efficiently handle complex, mixed variable types and nonlinear relationships typical of resource allocation problems of \ac{NS}. However, MILP and NLP are NP-Hard, thus, solutions cannot be efficiently computed for large-scale instances within polynomial time \cite{b15}. To address this, the problems are typically decomposed into simpler sub-problems, or the constraints are relaxed, allowing for the application of lower-complexity algorithms like heuristics and meta-heuristics to find near-optimal solutions. For example, in \cite{b16}, the authors use techniques like Big-M formulation and successive convex approximation to optimized RAN slicing for 5G networks. Their focus was on the joint allocation of power and \ac{RBs} to support eMBB, URLLC, and mMTC services under imperfect Channel State Information (CSI). However, the complexity of the allocation problem is recognized due to their combinatorial and non-convex nature, necessitating advanced computational techniques for practical implementation. 

Due to the inherent computational complexity of optimization-based models and their struggles with managing the increasingly complex and heterogeneous traffic
flows generated by 5G services, the research community is shifting focus to ML solutions in recent times. ML-based approaches are well-suited for handling large datasets and identifying complex patterns\cite{b17}, offering significant potential for the autonomous management of resources within the \ac{NS} framework.

\subsection{Machine Learning-based solutions to slicing problems}

In \cite{b18}, the authors propose a \ac{DRL}-based \ac{NS} method for Beyond 5G systems, aiming to optimize resource allocation for maximizing long-term throughput while adhering to diverse \ac{QoS} requirements. This approach strategically reduces the action space (actions equals RB allocations) by eliminating non-viable options, thereby accelerating the training process and enhancing policy quality. The authors in \cite{b19} utilize Deep Q-Networks (DQN) and Soft Actor Critic (SAC), to optimize slice allocation in 5G networks, showing improvements over heuristic methods in managing resources to meet \ac{QoS} demands. The study in \cite{b20} presents a digital-twin (DT) assisted framework for resource allocation in \ac{NS}, particularly targeting highly personalized services within Industry 4.0 and Beyond, using Distributed Deep Reinforcement Learning (DDRL). The research reveals that this method significantly enhances service equilibrium and accelerates algorithm convergence. Meanwhile, the authors in \cite{b21} introduces a federated learning-driven digital twin (FED-DT) framework for optimizing \ac{NS} in communication networks, using non-Euclidean graph representations and a novel Graph Lineformer Network (GLN) to collaboratively learn and predict \ac{QoS} metrics. Key findings highlight FED-DT ability to meet stringent \ac{QoS} requirements and improve network performance.

In \cite{b22}, the authors explored unsupervised deep learning(DL) for distributed \ac{SFC} embedding in large-scale network environments, employing a decentralized approach that utilizes clustering and unsupervised neural Distributed Deep Learning (DistrDL) networks for efficient embedding. The findings suggest that this method improves resource efficiency and reduces solver runtime significantly compared to centralized and heuristic methods. The study in \cite{b23} addresses the cold-start problem in \ac{NS} orchestration through a Federated Meta Reinforcement Learning (FedMRL) approach, contributing significantly by enhancing adaptation speed and reducing training costs while maintaining data privacy. The authors demonstrates that FedMRL outperforms conventional methods in terms of cost, latency, and convergence speed. \cite{b24} introduces a novel approach for selecting the optimal Deep Reinforcement Learning (DRL) algorithm for Virtual Network Embedding (VNE) challenges in beyond 5G networks, leveraging the Algorithm Selection (AS) paradigm to enhance resource allocation. Key findings demonstrate that this approach not only improves performance but also outperforms standalone algorithms. The authors in \cite{b25} focus on establishing a service-aware slice design policy that efficiently guides the implementation of end-to-end network slices tailored for specific 5G use cases, enhancing resource utilization and addressing fluctuated traffic demands using multiple-objective particle swarm optimization (MOPSO) framework. The findings show that the proposed \ac{NS} policy not only optimizes resource efficiency across various service scenarios but also manages to balance the diverse performance requirements of 5G services.
\begin{table*}[htbp]
\centering
\caption{Summary of key findings and limitations in recent machine learning-based solutions to Network Slicing}
\label{tab:ML_based NS solutions}
\begin{tabular}{|p{0.03\textwidth}|p{0.07\textwidth}|p{0.35\textwidth}|p{0.4\textwidth}|}
\hline
\textbf{Ref.} & \textbf{Framework} & \textbf{Key Findings} & \textbf{Limitations} \\
\hline
\cite{b18} & DRL & Optimise resource allocation for maximising long-term throughput, reduces action space. & Faces challenges in computational complexity and training. \\
\hline
\cite{b19} & DQN,SAC & Optimise slice allocation, improves resource management. & Still under development, needs real-world validation. \\
\hline
\cite{b20} & DT-DDRL & Boosts service equilibrium and speeds up algorithm convergence. & Managing complex service requirements is challenging. \\
\hline
\cite{b21} & FED-DT & meet stringent QoS requirements, enhanced network performance. & Complex to manage and train. \\
\hline
\cite{b22} & DistrDL & Improves resource efficiency and reduces solver runtime. & Faces challenges in managing large-scale embeddings and initial training. \\
\hline
\cite{b23} & FedMRL & Enhances adaptation speed and reduces training costs while maintaining
data privacy. & Complex to implement in dynamic environments. \\
\hline
\cite{b24} & AS-DRL & Improves performance and outperforms standalone algorithms. & Needs further enhancements for dynamic conditions. \\
\hline
\cite{b25} & MOPSO & Optimizes resource efficiency \& balances diverse performance requirements & Struggles with optimizing under variable conditions. \\
\hline
\end{tabular}
\end{table*}
\subsection{Limitations of the Machine Learning-based solutions}
The recent works \cite{b18} -\cite{b25} reviewed above offers significant advancements to 5G \ac{NS} problems focusing on dynamic resource allocation and enhanced network efficiency which are crucial for handling the complex, variable demands of industrial environments. Although the results are promising, several common limitations are evident:
\begin{enumerate}[label=\textit{\arabic*.}, itemsep=0pt, parsep=0pt, partopsep=0pt, topsep=0pt, leftmargin=0pt, labelwidth=*, labelindent=0pt, align=left]

\item \textit{Validation:} There is a recurring theme of insufficient real-world validation. The studies in \cite{b18} -\cite{b25} demonstrate promising results in simulated or controlled environments but lack comprehensive testing in real operational settings, which raises questions about their practical applicability.

\item \textit{Complexity:} the studies in \cite{b18,b21,b23,b25}, highlights the complexity of implementing these advanced algorithms. The complexity often relates to the computational demands and the sophistication required in setting up and tuning the models.

\item \textit{Scalability:} As networks grow in size and complexity, the scalability of proposed solutions becomes a challenge. \cite{b22,b23} emphasizes the struggle to maintain performance as the network scale increases.

\item \textit{Training and Convergence:} ML models, especially DL models, frequently encounter long training times and issues with convergence as highlighted in \cite{b18, b20, b23}. This is a significant hurdle in environments where network conditions change rapidly.

\item \textit{Handling Uncertainty:} the approaches reviewed, addressed the issue of fluctuating network conditions and traffic demands like in \cite{b25}. However, these models often assume a level of predictability or require adjustments that may not be feasible in practice, thus limiting their effectiveness under actual uncertain conditions.
\end{enumerate}

These limitations underscore the need for more robust, adaptable, and scalable solutions that can be effectively implemented and validated in real-world 5G network environments. Table \ref{tab:ML_based NS solutions} summarises the key findings in the recent works on \ac{ML}-based solutions for \ac{NS} and their limitations.

\section{Technical Scenarios}

\subsection{Requirements of smart factory use cases}
Currently, most industrial communication technologies are wired, which includes various dedicated Industrial Ethernet technologies and fieldbuses, used primarily for connecting sensors, actuators, and controllers\cite{b26}. However, the rise of Industry 4.0 and advancements in 5G technology may significantly transform this, as wireless connectivity is essential for the flexibility, mobility, and efficiency needed in future factories, potentially revolutionizing production and service processes.

In 3GPP TR 22.804\cite{b27}, five main categories of smart factory use cases has been identified: 
\begin{itemize}
\item\textit{Factory Automation}: leverages automated technologies like robotics and programmable logic to enhance manufacturing processes. Encompasses applications like motion control, control-to-control communications, mobile robots, and massive wireless sensor networks. 
\item\textit{Process Automation}: optimizes the production and management of materials, using sensors and controllers to improve efficiency and safety. It comprises applications such as closed-loop control, process monitoring, and plant asset management. \item\textit{HMIs and Production IT}: bridges personnel with production processes, incorporating devices from basic panels to AR/VR technologies. It includes applications such as mobile control panels with safety features and augmented reality.
\item\textit{Logistics and Warehousing}: manages the flow and storage of industrial materials and products, utilizing \ac{AGV} and systems for efficient intra-logistics and warehousing, which are essential for material handling and asset tracking. Applications here include control-to-control and mobile robots.
\item\textit{Monitoring and Maintenance}: focuses on the passive monitoring of processes and predictive maintenance to foresee and optimize maintenance needs using data analytics. Features applications like massive wireless sensor networks, remote access and maintenance.
\end{itemize}
Each of these applications have different requirements in terms of availability, latency, payload, and jitter. Hence, for effective deployment, different applications may require separate communication services which can be physical, logical, or virtual\cite{b27}. Table \ref{tab:requirements} summarises some key smart factory applications with their requirements.

\begin{table*}[htbp]
\centering
\caption{Smart factory applications and their requirements\cite{b27}}
\includegraphics[width=1.\textwidth]{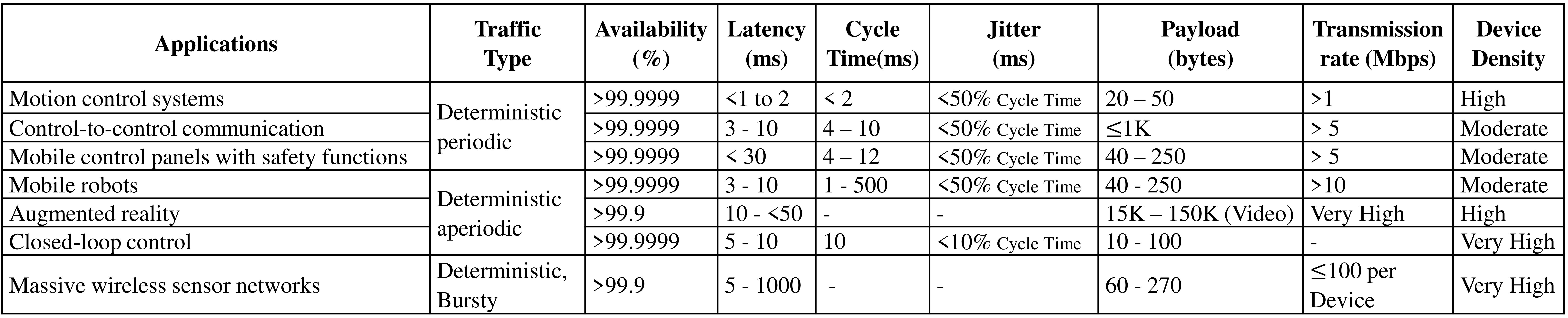}
\label{tab:requirements}
\end{table*}

\subsection{Smart Factory Network Elements}
Industrial 5G devices, which vary extensively in form and type, are adaptable for numerous applications. These devices can function either as standalone units or be integrated into other systems. In \cite{b28}, 5GACIA detailed these industrial 5G devices including;
\begin{enumerate}[label=\textit{\arabic*.}, itemsep=0pt, parsep=0pt, partopsep=0pt, topsep=0pt, leftmargin=0pt, labelwidth=*, labelindent=0pt, align=left]
    \item \textit{Programmable Logic Controllers:}
    PLCs serve as central units in industrial settings, managing operations of machinery via input from sensors and executing actions based on programmed settings. They are equipped with 5G radio interfaces for communication within control loops and with other PLCs, where timing is crucial.
    \item \textit{Sensors and Actuators:} These devices detect physical properties and perform actions based on control signals. In 5G settings, they connect via radio interfaces to PLCs and controllers, supporting real-time communication and high reliability. Low-power sensors focus on monitoring and conservation of energy, while 2D/3D sensors, often integrated with cameras and LIDAR, collect data for quality assurance and positioning, employing AI for analysis.

    \item \textit{Extended Reality and Human-Machine Interfaces:} Extended reality includes virtual and augmented reality technologies creating immersive digital environments. HMIs facilitate human interaction with machines, particularly in manufacturing, connecting to the 5G network and featuring devices such as video screens and cameras to provide vital operational information.
    
    \item \textit{Cellular Modems and Gateways:} These are crucial for linking devices with the cellular network, converting digital data for transmission and providing connectivity as routers. They enable Internet of Things (IoT) applications and remote communications, facilitating the efficient transfer of data between local and cloud-based systems.
    
    \item \textit{Cloud/Edge Computing Devices:} These devices handle data processing either at the source (edge computing) to reduce latency or remotely (cloud computing) to enhance scalability and efficiency. They connect to the industrial 5G network using compatible hardware to support various industrial applications.
\end{enumerate}

\subsection{Slice-Aware Radio Resource Management}
Slice-Aware \ac{RRM} refers to the methodologies and mechanisms employed to dynamically allocate and optimize radio resources such as spectrum, power, and spatial resources across different network slices\cite{b13}. Each slice caters to a specific set of service requirements, ranging from high-throughput data services to \ac{URLLC} necessary for real-time control systems in smart factories. The core idea behind Slice-aware \ac{RRM} is to provide a differentiated and customized network experience that aligns with the specific performance metrics of each slice.

Recent strategies for \ac{RRM} are characterized by innovative approaches that leverage advanced technologies such as \ac{AI} and \ac{ML}\cite{b29}, advanced MIMO and beamforming techniques\cite{b30}, and dynamic spectrum sharing\cite{b31}, to address the dynamic and complex requirements of network slicing, particularly in dense radio environments like smart factories. These strategies underscore a shift towards more intelligent, flexible, and efficient \ac{RRM} solutions to ensure that diverse service requirements of different slices are met with high precision. As detailed in Section III, many authors are currently focusing on \ac{ML} and \ac{AI} solutions to develop various slice aware \ac{RRM} frameworks.

\begin{figure}[htbp]
\centerline{\includegraphics[width=1 \linewidth]{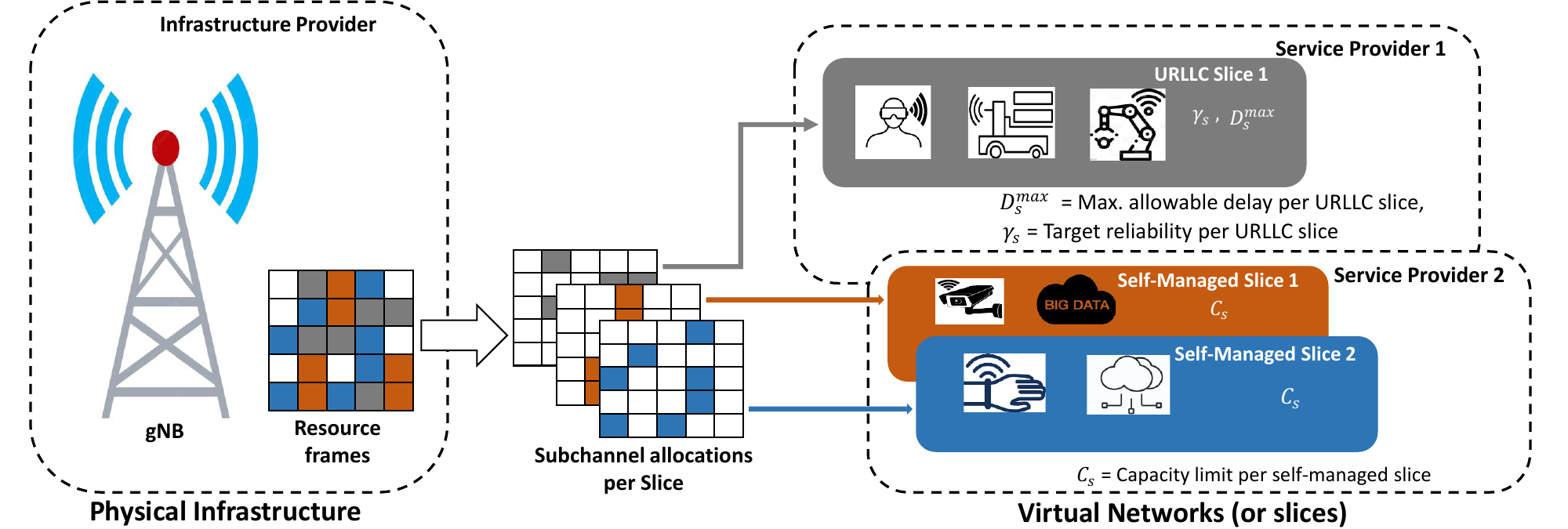}}
\caption{Sample system model of slice-aware resource allocation}
\label{fig:slice-aware_RRA}
\end{figure}

In \cref{fig:slice-aware_RRA}, a sample system model of a slice aware \ac{RRA} is shown. Here, we consider the downlink of a cellular network having a single \ac{BS} with one infrastructure provider(InP). Service providers (SPs) have contracts with the InP for one or more slices of the network with guaranteed end-to-end requirements for rate, latency, and reliability. Charges to the SPs are based on these criteria. 
The network features two main types of slices: \ac{URLLC} slices and Self-Managed (SM) slices. In the case of SM slices, SPs buy the capacity and oversee it independently, often leasing these slices to mobile Internet providers for a range of services focused mainly on capacity demands. Conversely, in URLLC slices, the InP offers SPs reliable, low-latency network slices typically used in applications like extended reality. The \ac{RRA} framework is tasked with determining the required subchannels and power allocation for network slicing, aiming to fulfil the requirements of both slice types while ensuring slice isolation. The resources a slice receives is determined by the requirements of the service it is associated with.

\subsection{Indoor-Factory Radio Propagation}
\ac{InF} environments such as manufacturing plants\cite{b32}\cite{b33}, feature high ceilings, expansive areas, and a significant presence of smooth metal surfaces that can cause specular reflections. These environments feature objects of diverse shapes, and irregular sizes, with machines as the primary cause of signal blockage rather than walls as is the case in indoor office environments\cite{b34}. Since the wireless channel is heavily impacted by the propagation environment, for precise modelling of the wireless channel in an \ac{InF} scenario, it is crucial to describe in detail the propagation environment. 

 The \ac{3GPP} specified the channel model for \ac{InF} in the \ac{3GPP} TR38.901 technical report\cite{b35}. This model takes into account environmental characteristics such as clutter density, antenna height, factory volume, and total surface area for the modelling of both large-scale and small-scale parameters. Also, to accommodate the mobility of terminals and positioning in industrial settings, dual mobility and absolute time of arrival have been incorporated into the model. 
 
 \ac{3GPP} TR38.901 specifies a \ac{LOS} and \ac{NLOS} path loss models. The \ac{NLOS} scenario incorporates the antenna height of the \ac{BS}, and the clutter density, with four sub-scenarios outlined; 1) Sparse clutter, low BS (SL), 2) Dense clutter, low BS (DL), 3) Sparse clutter, high BS (SH), and 4) Dense clutter, high BS (DH). The alpha-beta-gamma (ABG) model\cite{b34} was then used to specify the path-loss models. Additionally, models for channel parameters including the LOS probability, the root-mean-square (RMS) delay spread, and the angular spread, are provided in \ac{3GPP} TR38.901\cite{b35}.

\subsection{Traffic Characteristics}
Smart factory applications exhibit varied traffic patterns, from packets generated at regular intervals with low data volumes to irregular packets that sometimes involve substantial data volumes. This section outlines the types of data traffic typical in industrial scenarios as defined by \ac{5GACIA} \cite{b36}.

\begin{enumerate}[label=\textit{\arabic*.}, itemsep=0pt, parsep=0pt, partopsep=0pt, topsep=0pt, leftmargin=0pt, labelwidth=*, labelindent=0pt, align=left]
    
    \item \textit{Deterministic periodic traffic}  is characterized by data transmissions occurring at fixed intervals, with the data expected to arrive with uniform latency during these intervals. The acceptable levels of latency and jitter are dependent on the specific application. For example, packets from sensors to controllers are sent periodically among synchronized, time-sensitive applications. This covers situations such as emergency stop signals from handheld controllers to a machine \ac{PLC}. To specify deterministic periodic traffic, the packet size, and the transfer interval are used.
    \item \textit{Deterministic aperiodic traffic} involves data that are sent regularly but not at predetermined interval, thus classified as aperiodic. This traffic type is defined by their packet size and an average transfer interval. For example, sensor-to-controller communications may occur due to changes in status or values within non-synchronized, time-sensitive applications, such as an optical sensor relaying position data to a conveyor belt PLC.
    \item \textit{Non-deterministic traffic and burst traffic.} Non-deterministic traffic may be either periodic or aperiodic. This traffic type is characterized by the absence of strict timing requirements for message delivery at the destination. The essential requirement is that messages arrive accurately and in the correct sequence. Burst traffic, on the other hand, is characterized by consecutive sequences of messages sent in a "burst," which can be used to transmit images for example. Parameters such as average data rate and peak data rate are used to describe non-deterministic and burst traffic. Drones sending real-time video feeds to a monitoring station, and test report uploads, are examples of applications that generate this type of traffic.
\end{enumerate}

\subsection{Enabling Technologies}
Implementing end-to-end network slicing for smart factories involves a complex integration of several advanced technologies to create a tailored, flexible, and efficient network architecture. Technologies such as \textit{Network Functions Virtualization (NFV)} and \textit{Software-Defined Networking (SDN)} abstract and manage hardware and network functions to rapidly deploy and dynamically scale network capabilities as needed\cite{b13}\cite{b37}. \textit{Cloud-native technologies}, such as microservices and containerization, along with \textit{Edge Computing}, enhance agility and reduce latency, which are essential in manufacturing environments with stringent real-time requirements\cite{b38}. Additionally, \textit{AI/ML} optimize network operations by predicting and adapting to network conditions in real time\cite{b34}, while \textit{orchestration systems} coordinate all resources to ensure each network slice meets its specific needs\cite{b39}. \textit{Virtual Machines (VMs)} and \textit{Containers} further support these capabilities by allowing isolated and rapid scaling of network functions, crucial for maintaining consistent performance and flexibility in smart manufacturing settings\cite{b13}.

\subsection{Standardization Efforts}
3GPP is a key player in the standardization of 5G \ac{NS}, with several working groups such as SA1 defining use cases and requirements, SA2 designing the system architecture, and SA3 focusing on security features. Additional groups like SA5 manage network slices in coordination with other Standard Development Organizations (SDOs), while RAN1/2/3 develop RAN slicing features, collectively contributing to a robust end-to-end network slice framework.\\
\indent 3GPP \ac{NS} standardisation efforts from Release (Rel-)15 to 18 \cite{b40} is highlighted as follows: \textit{Rel-15} introduced fundamental features for network slice lifecycle management, encompassing the introduction of network slice concepts, provisioning management, performance monitoring, and fault supervision. Subsequently, \textit{Rel-16} brought enhancements including Service Level Agreement (SLA) attributes and the introduction of closed-loop automation for better management adaptability. In \textit{Rel-17}, the focus shifted to the management of non-public networks through network slicing and the implementation of a closed-loop assurance mechanism tailored to meet diverse SLA demands. This release also advanced the discussion on \ac{NS} with specific attention to Energy Efficiency (EE) Key Performance Indicators (KPIs) and SLA improvements. \textit{Rel-18} aim to further enhance provisioning efficiency using rules-based frameworks and asynchronous operations. Initiatives in this release also cover supporting energy utilities service providers, making network slices more transparent to customers, and managing slices through intent-driven approaches. The ongoing discussions in \textit{Rel-19}, identified \ac{NS} as one of the 15 "miscellaneous" potential topics for inclusion in the release. Moreover, since Rel-16, there has been a consistent enhancement in network slicing-related performance measurements and KPIs, showcasing continual improvements in network slice management and operational efficiency across successive releases.

Other standardisation bodies like the 5G Automotive Association (5GAA) collaborate with automotive and telecommunications sectors to develop end-to-end solutions for future mobility
and transportation services, while the Industrial Internet Consortium (IIC) and the German Electrical and Electronic Manufacturers' Association (ZVEI) push forward 5G smart manufacturing. Meanwhile, the European Telecommunications Standards Institute (ETSI) and the International Telecommunication Union - Telecommunication (ITU-T) are enhancing 5G network slicing and setting standards for deployment, configuration, and orchestration, addressing a range of service requirements and infrastructure architectures. And NGMN continue to develop, refine, and disseminate the requirements and architectural framework for 5G network slicing.

\section{Network Slicing Challenges in Smart Factories: Research Frontiers}
In the dynamic and complex environment of smart factories, implementing 5G network slicing presents several unique and significant open research challenges that need to be addressed to fully harness the potential of this technology:

\begin{enumerate}[label=\textit{\arabic*.}, itemsep=0pt, parsep=0pt, partopsep=0pt, topsep=0pt, leftmargin=0pt, labelwidth=*, labelindent=0pt, align=left]
\item \textit{Time-Sensitive Networking:} Time-Sensitive Networks (TSN) represents a crucial component of Industry 4.0, focusing on the precise timing and synchronization needed for critical industrial applications. TSN is currently defined within the IEEE 802.1 standards for wired networks, but there is ongoing momentum to integrate these capabilities into 3GPP cellular wireless technologies. The goal is to leverage the ubiquity and flexibility of wireless networks while maintaining the stringent requirements of time sensitivity and reliability.
\item \textit{Terminal Mobility:} effectively managing mobility to ensure seamless handovers without communication disruption, particularly when integrated with edge computing is still a research challenge. This necessitates sophisticated mechanisms to dynamically adjust network parameters and resources, ensuring that data integrity and processing continuity are preserved during handover events.
\item \textit{Strict QoS Requirements:} Addressing the industrial-grade \ac{QoS} requirements, such as stringent end-to-end latency and determinism for smart factories, remains an active area of research. These requirements stem from the need to support diverse critical applications.
\item \textit{Integration with Existing Wired Technologies:} Seamless operation in smart factories requires the integration of 5G with existing wired connectivity technologies. This integration is vital for maintaining continuous operations across machinery and production lines, presenting a complex challenge in system design and interoperability.
\item \textit{Radio Propagation:} the factory environment, often filled with metallic objects and electrical interference from various industrial activities, poses specific research challenges for indoor-factory radio propagation.
\item \textit{Slice-Aware Resource Management:} developing intelligent and autonomous RRM strategies that can proactively allocate resources based on anticipated network conditions in dynamic and demanding smart factory environments is still an open research challenge.
\end{enumerate}

\section{Conclusion and Future Expectation}
This paper explores network slicing to enhance factories of the future, emphasizing its importance in advancing Industry 4.0 by improving operational efficiency and offering tailored network solutions. We discussed the technology, recent advancements, and the integration of AI/ML to meet the dynamic needs of modern manufacturing. However, deploying network slicing in smart factory networks presents significant challenges, such as complexity in dense radio environments and the need for effective slice management to maintain reliability and performance under varying conditions.

While significant progress has been made, the full potential of network slicing for smart factories remains to be realized. Ongoing research, innovation, and collaboration are crucial to address current challenges and harness the transformative power of network slicing for manufacturing. Looking forward, the development of network slicing will greatly benefit from 6G technologies, offering better connectivity, energy efficiency, and lower latency. Future research focusing on advanced predictive management and adaptive resource allocation is essential. Moreover, extensive real-world testing and cross-sector collaboration are vital to refine these technologies and meet the stringent demands of smart factory networks.

\section*{Acknowledgment}

This work is supported by the 5GSmartFact project, which has received funding from the European Union's H2020 research and innovation program under the Marie Skłodowska-Curie grant ID 956670. Also by the project 6-SENSES grant PID2022-138648OB-I00 funded by MCIN/AEI/10.13039/501100011033 and by FEDER-UE, ERDF-EU \textit{A way of making Europe}, and the grants 22CO1/008248 and 2021 SGR 01033 (AGAUR, Generalitat de Catalunya).

\vspace{12pt}


\begin{thebibliography}{00}

\bibitem{b1} A. Radziwon, A. Bilberg, M. Bogers, and E. S. Madsen, "The smart factory: exploring adaptive and flexible manufacturing solutions," Procedia Engineering, vol. 69, pp. 1184-1190, 2014.

\bibitem{b2} M. Soori, B. Arezoo, and R. Dastres, "Internet of things for smart factories in industry 4.0, a review," Internet of Things and Cyber-Physical Systems, 2023.

\bibitem{b3} D. Mourtzis, J. Angelopoulos, and N. Panopoulos, "Smart manufacturing and tactile internet based on 5G in industry 4.0: Challenges, applications and new trends," Electronics, vol. 10, no. 24, p. 3175, 2021. 

\bibitem{b4} X. Foukas, G. Patounas, A. Elmokashfi, and M. K. Marina, "Network slicing in 5G: Survey and challenges," IEEE Communications Magazine, vol. 55, no. 5, pp. 94-100, 2017.

\bibitem{b5} T-Specification, ‘‘Study on architecture for next generation system,
release 14,’’ 3GPP, Sophia Antipolis, France, Tech. Rep. TR 23.799, 2016.

\bibitem{b6} A. E. Kalør, R. Guillaume, J. J. Nielsen, A. Mueller, and P. Popovski, "Network slicing in industry 4.0 applications: Abstraction methods and end-to-end analysis," IEEE Transactions on Industrial Informatics, vol. 14, no. 12, pp. 5419-5427, 2018

\bibitem{b7} 3GPP TS 23.501, "System Architecture for the 5G System," [Online]. Available: \url{https://www.3gpp.org/ftp/Specs/archive/23_series/23.501/}.

\bibitem{b8} 3GPP TS 22.261 "Service requirements for next generation new services and markets". [Online]. Available: \url{https://www.3gpp.org/ftp/Specs/archive/22_series/22.261/}.

\bibitem{b9} NGMN Alliance, "NGMN 5G White Paper," February 2015. [Online]. Available: \url{https://www.ngmn.org/wp-content/uploads/NGMN_5G_White_Paper_V1_0.pdf}.

\bibitem{b10} Description of Network Slicing Concept, NGMN 5G P1 Requirements and Architecture, Work Stream End-to-End Architecture, Version 1.0, NGMN Alliance, Frankfurt, Germany, Jan. 2016.

\bibitem{b11} S. Wijethilaka and M. Liyanage, "Survey on network slicing for Internet of Things realization in 5G networks," IEEE Communications Surveys and Tutorials, vol. 23, no. 2, pp. 957-994, 2021.

\bibitem{b12} 3GPP TS 22.530 "Management and orchestration; Concepts, use cases and requirements". [Online]. Available: \url{https://www.3gpp.org/ftp/Specs/archive/28_series/28.530/} 

\bibitem{b13} Y. Wu, H. N. Dai, H. Wang, Z. Xiong, and S. Guo, "A survey of intelligent network slicing management for industrial IoT: Integrated approaches for smart transportation, smart energy, and smart factory," IEEE Communications Surveys and Tutorials, vol. 24, no. 2, pp. 1175-1211, 2022.

\bibitem{b14} T. Umagiliya, S. Wijethilaka, C. De Alwis, P. Porambage, and M. Liyanage, "Network slicing strategies for smart industry applications," in 2021 IEEE Conference on Standards for Communications and Networking (CSCN), 2021, pp. 30-35.

\bibitem{b15} H. P. Phyu, D. Naboulsi, and R. Stanica, "Machine learning in network slicing—a survey," IEEE Access, 2023.

\bibitem{b16} P. K. Korrai, E. Lagunas, A. Bandi, S. K. Sharma, and S. Chatzinotas, "Joint power and resource block allocation for mixed-numerology-based 5G downlink under imperfect CSI," in IEEE Open Journal of the Communications Society, vol. 1, pp. 1583-1601, 2020.

\bibitem{b17} A. Zappone, M. Di Renzo, and M. Debbah, "Wireless networks design in the era of deep learning: Model-based, AI-based, or both?," IEEE Transactions on Communications, vol. 67, no. 10, pp. 7331-7376, Oct. 2019

\bibitem{b18} K. Suh, S. Kim, Y. Ahn, S. Kim, H. Ju, and B. Shim, "Deep reinforcement learning-based network slicing for beyond 5G," IEEE Access, vol. 10, pp. 7384-7395, 2022.

\bibitem{b19} I. Amonarriz-Pagola and J. A. Fernandez-Carrasco, "A Reinforcement Learning Approach for Network Slicing in 5G Networks," 2023 JNIC Cybersecurity Conference (JNIC), IEEE, 2023.

\bibitem{b20} L. Tang, Y. Du, Q. Liu, J. Li, S. Li, and Q. Chen, "Digital twin assisted resource allocation for network slicing in industry 4.0 and beyond using distributed deep reinforcement learning," IEEE Internet of Things Journal, 2023.

\bibitem{b21} M. Abdel-Basset et al., "Digital Twin for Optimization of Slicing-enabled Communication Networks: A Federated Graph Learning Approach," IEEE Communications Magazine, 2023.

\bibitem{b22} P. Rodis and P. Papadimitriou, "Unsupervised Deep Learning for Distributed Service Function Chain Embedding," IEEE Access, 2023.

\bibitem{b23} T. Dong, Q. Qi, J. Wang, Z. Zhuang, H. Sun, J. Liao, and Z. Han, "Standing on the Shoulders of Giants: Cross-Slice Federated Meta Learning for Resource Orchestration to Cold-Start Slice," IEEE/ACM Transactions on Networking, 2022.

\bibitem{b24} A. Bouroudi, A. Outtagarts, and Y. Hadjadj-Aoul, "Dynamic Machine Learning Algorithm Selection For Network Slicing in Beyond 5G Networks," 2023 IEEE 9th International Conference on Network Softwarization (NetSoft), IEEE, 2023.

\bibitem{b25} R. Wang, A.-H. Aghvami, and V. Friderikos, "Service-aware design policy of end-to-end network slicing for 5G use cases," IEEE Transactions on Network and Service Management, vol. 19, no. 2, pp. 962-975, 2022.

\bibitem{b26} R. Zurawski, "Industrial communication technology handbook", second edition, CRC Press, September 2017.

\bibitem{b27} 3GPP, "Study on Communication for Automation in Vertical Domains (Release 16)," TR 22.804, Sophia Antipolis, France, 2020. [Online]. Available: \url{https://www.3gpp.org/ftp/Specs/archive/22_series/22.804/}

\bibitem{b28} 5G-ACIA, "Industrial 5G Devices: Architecture and Capabilities," 5G Alliance for Connected Industries and Automation. [Online]. Available: \textit{https://5g-acia.org/whitepapers/industrial-5g-devices-architecture-and-capabilities/}.

\bibitem{b29} X. Shen, J. Gao, W. Wu, K. Lyu, M. Li, W. Zhuang, et al., "AI-assisted network-slicing based next-generation wireless networks," IEEE Open Journal of Vehicular Technology, vol. 1, pp. 45-66, 2020.

\bibitem{b30} W. Chen, X. Lin, J. Lee, A. Toskala, S. Sun, C. F. Chiasserini, and L. Liu, "5G-advanced toward 6G: Past, present, and future," IEEE Journal on Selected Areas in Communications, vol. 41, no. 6, pp. 1592-1619, 2023.

\bibitem{b31} M. Abdelraheem and M. M. Abdellatif, "A stochastic spectrum trading and resource allocation framework for opportunistic dynamic spectrum access networks," IEEE Access, vol. 10, pp. 73774-73785, 2022.

\bibitem{b32} T. Jiang, L. Tian, J. Zhang, Y. Zheng, Q. Wang, and J. Dou, “The impact of antenna height on the channel model in indoor industrial scenario,” in 2020 IEEE/CIC International Conference on Communications in China (ICCC Workshops). IEEE, 2020, pp. 1–6.

\bibitem{b33} 5G-ACIA, "Industrial 5G Devices: Architecture and Capabilities," 5G Alliance for Connected Industries and Automation. [Online]. Available: \url{https://5g-acia.org/whitepapers/industrial-5g-devices-architecture-and-capabilities/}.

\bibitem{b34} Jiang, T., Zhang, J., Tang, P., Tian, L., Zheng, Y., Dou, J., Asplund, H., Raschkowski, L., D’Errico, R., and Jämsä, T. (2021). 3GPP standardized 5G channel model for IIoT scenarios: A survey. IEEE Internet of Things Journal, 8(11), 8799-8815.

\bibitem{b35} 3GPP, “Study on channel model for frequencies from 0.5 to 100 GHz, V16.0.0,” 3GPP, Sophia Antipolis, France, Rep. TR 38.901, Oct. 2019. [Online]. Available: \url{https://www.3gpp.org/ftp/Specs/archive/38_series/38.901/}

\bibitem{b36} 5G-ACIA, "A 5G Traffic Model for Industrial Use Cases," 5G Alliance for Connected Industries and Automation. [Online]. Available: \url{https://5g-acia.org/whitepapers/a-5g-traffic-model-for-industrial-use-cases/}.

\bibitem{b37} I. Afolabi, T. Taleb, K. Samdanis, A. Ksentini, and H. Flinck, "Network slicing and softwarization: A survey on principles, enabling technologies, and solutions," IEEE Communications Surveys and Tutorials, vol. 20, no. 3, pp. 2429-2453, 2018.

\bibitem{b38} S. D. A. Shah, M. A. Gregory, and S. Li, "Cloud-native network slicing using software defined networking based multi-access edge computing: A survey," IEEE Access, vol. 9, pp. 10903-10924, 2021.

\bibitem{b39} Barakabitze, A. A., Ahmad, A., Mijumbi, R., and Hines, A. (2020). 5G network slicing using SDN and NFV: A survey of taxonomy, architectures and future challenges. Computer Networks, 167, 106984.

\bibitem{b40} 3GPP "5G Network slice management," July, 2023. [Online]. Available: \url{https://www.3gpp.org/technologies/slice-management}

\end{thebibliography}
\end{document}